\documentclass[aps,preprint,prd,epsfig]{revtex4}
\usepackage{amssymb}
\usepackage{amsmath}
\usepackage{graphicx}
\usepackage{fancyhdr}
\newcommand{\bea}{\begin{eqnarray}}
\newcommand{\eea}{\end{eqnarray}}

\begin{document}

\title{ Simple Realization of the Inverse Seesaw Mechanism}

\author{A. G. Dias$^{1}$,  C. A. de S. Pires$^{2}$, P. S. Rodrigues da Silva$^{2}$, A. Sampieri$^{2}$}

\affiliation{\vspace{0.8cm}
\\
 $^{1}$ Centro de Ci\^encias Naturais e Humanas, Universidade Federal do ABC,  Santo Andr\'e-SP, 09210-170, BraZil.\vspace{0.4cm}
 \\
 $^{2}$ Departamento de
F\'{\i}sica, Universidade Federal da Para\'\i ba, Caixa Postal 5008, 58051-970,
Jo\~ao Pessoa, PB, Brazil
 }

\date{\today}

\begin{abstract}
Differently from the canonical seesaw mechanism, which is grounded in grand unified theories, the inverse seesaw mechanism lacks a special framework that realizes it naturally. In this work we advocate that the 3-3-1 model with right-handed neutrinos has such an appropriate framework to accommodate the inverse seesaw mechanism.  We also discuss  the smallness of the lepton number violating mass  and estimate the branching ratio for the rare lepton flavor violation process $\mu \rightarrow e\gamma$.
\\
%PACS: 14.60.St; 14.60.Pq; 12.60.Cn; 12.60.Fr.
%
\end{abstract}

\maketitle

\section{Introduction}
\label{sec1}
Although experiments in neutrino oscillations   have reported that neutrinos are  light particles mixed in an  unusual  way~\cite{experiments},
\begin{eqnarray}
&& \Delta m^2_{21}= (7.59 \pm 0.21) \times10^{-5}\mbox{ eV}^2\,,\,  \Delta m^2_{31}= (2.43 \pm 0.13) \times 10^{-3}\mbox{eV}^2,\nonumber \\
&& \sin^2(2\theta_{12})=0.861^{+0.026}_{-0.022} \,\,\,,\,\,\, \sin^2(2\theta_{23})>0.92\,\,\,,\,\,\,\sin^2(2\theta_{13})= 0.092 \pm 0.016,
\label{currentneutrinodata}
\end{eqnarray}
from the theoretical side we still miss a definitive understanding of the smallness of the  neutrino masses and of the profile of their mixing.

Seesaw mechanisms~\cite{seesawI,seesawII,seesawIII} are considered the most elegant way of explaining the smallness of the neutrino masses.  Their essence  lies in the fact that lepton number must be explicitly violated at a high energy scale.  As a result,  left-handed neutrinos gain small masses through the formula  $m_\nu = \frac{v^2_w}{\Lambda}$, where $v_w$   is the electroweak scale and $\Lambda$ is associated to the lepton number violation scale.  In the  seesaw mechanisms $\Lambda$ is generally related to some grand unified theory scale. In this way, for  $\Lambda=10^{14}$~GeV, we get neutrino masses at sub-eV scale. In spite of providing an interesting explanation for the smallness of the neutrino masses, such mechanisms are not phenomenologically testable because the  new physics engendered by them will manifest at $10^{14}$~GeV scale which is completely out of the range of the current and next accelerator experiments.

A radically different  realization of the seesaw mechanism is the so-called inverse seesaw mechanism (ISS)~\cite{inverseseesaw}, where small neutrino masses arise as a result of new physics at TeV scale which may be probed at the Large Hadron Collider (LHC) experiments. According to the original idea, the implementation of the ISS mechanism requires the addition of three right-handed neutrinos $N_{iR}$ and three extra standard model singlet neutral fermions, $S_{iL}$, to the three active neutrinos, $\nu_{i_L}$, with $i=1,2,3$.  The mechanism arises when we make use of extra symmetries in order to allow that these nine neutrinos  develop exactly  the following bilinear  terms,
\begin{equation}
{\cal L}= -\bar \nu_L  m_D N_R  - \bar S_L  M N_R -  \frac{1}{2}\bar S_L  \mu S_L^C + H.c,
\label{massterms}
\end{equation}
where $m_D$, $M$ and $\mu$  are generic $3\times 3 $ complex mass matrices. These terms  can be arranged  in the following  $9\times 9$ neutrino mass matrix  in the basis {\bf $(\nu_L\,,\,N_L^C\,,\,S_L)$},
\begin{equation}
M_\nu=
\begin{pmatrix}
0 & m^T_D & 0 \\
m_D& 0 & M^T\\
0 & M & \mu
\end{pmatrix}.
\label{ISSmatrix}
\end{equation}
On considering the hierarchy  $\mu<< m_D<<M$, the diagonalization of this $9 \times 9$ mass matrix provides  the following effective neutrino mass matrix for the standard neutrinos:
\begin{equation}
m_\nu = m_D^T (M^T)^{-1}\mu M^{-1} m_D.
\label{inverseseesaw}
\end{equation}
The double suppression by the mass scale connected with $M$ turns it possible to have such scale much below than that one involved in the canonical seesaw mechanism. It happens that standard neutrinos with mass at sub-eV scale are obtained for $m_D$ at electroweak scale, $M$ at TeV scale and $\mu$ at keV scale. In this case all the six RH neutrinos may develop masses around TeV scale and their mixing with the standard neutrinos is modulated by the ratio $m_DM^{-1}$. The core of the ISS is that the smallness of the neutrino masses is guaranteed by assuming that the $\mu$ scale is small and, in order to bring the RH neutrino masses down to TeV scale, it has to be at the keV scale~\cite{explanation1,explanation2}.

Differently from the canonical seesaw mechanism that finds its natural place in grand unified theories, the ISS mechanism still lacks a special framework where the six new neutrinos could be part of some underlying particle content  and naturally provide the mass terms  in Eq.~(\ref{massterms}). In this work we show that the $SU(3)_C \times SU(3)_L \times U(1)_N$ model with right-handed neutrinos (331RHN for short)~\cite{331original} has the appropriate framework to accommodate the  ISS mechanism.  This is so because this is a model which  may manifest at TeV scale and possesses in its matter content the new six right-handed neutrinos required by the mechanism and easily provides the mass terms in Eq.~(\ref{massterms}). In addition we develop an explanation for the smallness of the $\mu$ parameter and compute the branching ratio for the rare  lepton flavor violation process $\mu \rightarrow e\gamma$, for which stringent bounds are expected to emerge in future neutrino experiments~\cite{futureexperiments}.

In what follows we implement the ISS mechanism in the 331RHN and, then, we develop a suitable mechanism to explain the smallness of the $\mu$ parameter.

\section{ISS in the 3-3-1 model with right-handed neutrinos}
\label{sec3}

We consider the 331RHN~\cite{331original}  whose leptonic sector is composed by,
\begin{eqnarray}
& & f_{aL} = \left (
\begin{array}{l}
\nu_{aL} \,\, e_{aL} \,\, \nu^{C}_{aL}
\end{array}
\right )^T\sim(3\,,\,-1/3)\,,\nonumber\\
& & e_{a_R}\,\sim(1,-1)\,,\,\,\,\,\,\,\,\,\,N_{a_R}\sim(1,0),
\end{eqnarray}
where $a=1,2,3$,  and the numbers between parentheses refer to the $SU(3)_L$, $U(1)_N$ transformation properties. In this way we have the minimum  matter content needed in the ISS, i. e., nine neutral chiral leptons.

In order to generate mass to all fermions consistently, and also leaving only the electromagnetic symmetry group $U(1)_{em}$ explicitly realized, we take into account the following three scalar triplets
\begin{eqnarray}
& & \eta=
\left (\begin{array}{l}
\eta^{0} \,\,
\eta^{-} \,\,
\eta'^{0}
\end{array}
\right )^T\sim(3\,,\,-1/3)\,,\nonumber\\
& & \chi=
\left (\begin{array}{l}
\chi^{0} \,\,
\chi^{-} \,\,
\chi'^{0}
\end{array}
\right )^T\sim(3\,,\,-1/3)\,,\nonumber\\
& & \rho=
\left (\begin{array}{l}
\rho^{+} \,\,
\rho^{0} \,\,
\rho'^{+}
\end{array}
\right )^T\sim(3\,,\,2/3).
\label{scalarcontent}
\end{eqnarray}

The relevant Yukawa Lagrangian for the lepton sector that yields the ISS mechanism for the neutrinos is composed by the following summation of terms,
\begin{equation}
{\cal L}^Y_{\mbox{ISS}}=G_{ab}\epsilon_{ijk}\bar{L^C_{a_i}}\rho^*_j L_{b_k} + G^{\prime}_{ab}\bar L_a \chi N_{b_R} + \frac{1}{2} \bar N^C_R \mu N_R + H.c.
\label{yukawainteractions}
\end{equation}

We assume that  the fields $\eta^0$, $\rho^0$ and   $\chi^{\prime 0}$ develop vacuum expectation value (VEV) according to
\begin{eqnarray}
 \langle\eta^0\rangle=\frac{v_{\eta}}{\sqrt2},\,\,\,\,\,\,
\langle\rho^0\rangle=\frac{v_{\rho}}{\sqrt2},\label{veta}\,\,\,\,\,\,
 \langle\chi^{\prime }\rangle=\frac{v_{\chi^{\prime }}}{\sqrt2}.
\label{vrho}
\end{eqnarray}

With this set of VEVs, the Lagrangian above yields the following neutrino mass terms,
\begin{equation}
{\cal L}_{\mbox{mass}}= \bar \nu_L m_D \nu_R+  \bar \nu^{ C}_L M N_R + \frac{1}{2} \bar N^C_R \mu  N_R + H.c.
\label{massterms}
\end{equation}

In the basis $S_L =(\nu_L\,,\, \nu^{C}_L\,,\,N^C_L)$, the mass terms above can be cast in the following manner,
\begin{equation}
{\cal L}_{\mbox{mass}}=\frac{1}{2} \bar S^C_L M_\nu  S_L + H.c.,
\label{massterm}
\end{equation}
with the mass matrix $M_\nu$ having the texture,
\begin{equation}
M_\nu=
\begin{pmatrix}
0 & m^T_D & 0 \\
m_D& 0 & M^T\\
0 & M & \mu
\end{pmatrix}.
\label{massmatrix331}
\end{equation}
where the $3\times3$ matrices are defined as
\begin{eqnarray}
& & M_{ab}=G^{\prime}_{ab}\frac{v_{\chi^{\prime}}}{\sqrt2}\label{mM}\\
& & m_{Dab}=G_{ab}\frac{v_\rho}{\sqrt2}
\label{mmatrix3x3}
\end{eqnarray}
with  $M_{ab}$ and $m_{D_{ab}}$ being Dirac mass matrices, with this last one being anti-symmetric. The mass matrix in Eq.~(\ref{massmatrix331}) is characteristic of the ISS mechanism. We would like to call the attention to the fact  that the two energy scales related with the model's gauge symmetry breakdown appear in the mass matrix. Namely,  $v_{\chi^{\prime}}$ in $M_{ab}$ is connected with $SU(3)_L \otimes U(1)_N/SU(2)_L\otimes U(1)_Y$ and could be expected to be at the TeV scale leading to observable effects at the LHC, while $v_{\rho}$ in $m_{Dab}$ is connected with the electroweak standard model symmetry breakdown scale. Unfortunately the third  scale of energy, $\mu$,  characteristic of the mechanism is not a natural outcome of the 331RHN model. For the smallness of $\mu$, we provide, in the next section, an explanation inspired in that one formulated in Ref.~\cite{explanation1}.

In order to see how $M_\nu$ in Eq.~(\ref{massmatrix331}) can lead to eigenvalues at the eV scale it is useful to define the matrices,
\begin{equation}
{\cal M}_{D_{6\times3}}=
\begin{pmatrix}
m_{D_{3\times3}} \\
0_{3\times3}
\end{pmatrix},\,\,\,\,\, {\cal M}_{R_{6\times6}}=
\begin{pmatrix}
0_{3\times3} & M^T_{3\times3}  \\
M_{3\times3} &  \mu_{3\times3}
\end{pmatrix},
\label{definition}
\end{equation}
so that we have the following matrix with blocks, where ${\cal M}_R$ is supposed invertible matrix,
\begin{equation}
M_{\nu_{9\times9}}=
\begin{pmatrix}
0_{3\times3} & {\cal M}^T_{D_{3\times6}}  \\
{\cal M}_{D_{6\times3}} & {\cal M}_{R_{6\times6}}
\end{pmatrix}
\label{redefinedmassmatrix331}
\end{equation}
This last matrix can be diagonalized by means of procedures involving block matrices which is presented in Refs.~\cite{schechter-vale82}, \cite{hlr2011}. Following these Refs., we define a diagonalizing matrix, $W$, such that,
\begin{equation}
W^TM_\nu W=
\begin{pmatrix}
m_{light_{_{3\times3}}} & 0_{3\times6}  \\
0_{6\times3} & m_{heavy_{_{6\times6}}}
\end{pmatrix}.
\label{dmatrix}
\end{equation}
In this way, the $W$ matrix has the following form,
\begin{equation}
W=
\begin{pmatrix}
(\sqrt{1+FF^\dagger})_{3\times3} & F_{3\times6}  \\
F^\dagger_{6\times3} & (\sqrt{1+F^\dagger F})_{6\times6}
\end{pmatrix},
\label{Wm}
\end{equation}
where it is understood that,
\begin{equation}
\sqrt{1+FF^\dagger} \equiv 1-\frac{1}{2}FF^\dagger-\frac{1}{8}FF^\dagger FF^\dagger+...
\end{equation}
Under the assumption that $F$ is given as a power series in ${\cal M}_R^{-1}$,
\begin{equation}
F= F_1+F_2+...,
\end{equation}
\begin{equation}
F_i \sim ({\cal M}_R^{-1})^i,
\end{equation}
the eigenvalues of ${\cal M}_R$ are supposed to be larger than the entries of ${ \cal M}_D$. This is justified observing that the entries of $M$ in ${\cal M}_R$ are of order $v_{\chi^\prime}$.  Then, Eqs.~(\ref{dmatrix}) and (\ref{Wm}) allow to determine the blocks $m_{light}$ and $m_{heavy}$ order by order in ${\cal M}_R^{-1}$. At lowest order,
\begin{eqnarray}
&& F \approx  F_1 = ({\cal M}^T_D{\cal M}_R^{-1})^*\,  ,\nonumber \\
&&m_{light} \approx -{\cal M}^T_D {\cal M}^{-1}_R  {\cal M}_D\, ,\nonumber \\
&&m_{heavy} \approx {\cal M}_R.
\label{mlmh}
\end{eqnarray}
In general grounds these results are identical to those obtained from usual seesaw mechanism. What turns it different is the texture  of the matrices ${\cal M}_D$ in Eq. (\ref{definition}) and,
\begin{equation}
{\cal M}_R^{-1}=
\begin{pmatrix}
-M^{-1}\mu (M^T)^{-1} & M^{-1}  \\
(M^T)^{-1} & 0
\end{pmatrix},
\label{MRinv}
\end{equation}
which leads to the ISS form for the light Majorana neutrino mass matrix,
\begin{eqnarray}
m_{light} = m^T_D M^{-1}\mu (M^T)^{-1}m_D.
\label{mlf}
\end{eqnarray}
The minimal model we are developing here has the peculiar characteristic that $m_D$  is an anti-symmetric matrix. As the three active standard neutrinos masses correspond to the  eigenvalues of  Eq.~(\ref{mlf}), there is a prediction that one of them is massless.

A departure from a scenario involving just three active neutrinos, where their mixing is described by an unitary PMNS matrix, is observed in neutrinos mixing relying on the ISS. It happens that the largest energy scale figuring  in Eq.~(\ref{definition}) is $M\sim v_\chi^\prime$,  supposedly of 1~TeV order. As a consequence description of oscillation involving three active neutrinos will be attached with nonunitary effects modulated by  the ratio $v_W^2/v_\chi^{\prime 2}$. Such effects manifest experimentally through neutrino disappearing in disacordance from what is expected when considering unitarity in oscillation phenomena involving the three known neutrinos.

It is worthwhile to review how the nonunitarity aspect is quantified in the ISS mechanism. For obtaining the complete mass eigenstates the matrix in Eq. (\ref{dmatrix}) has still to be transformed to a diagonal form by means of,
\begin{equation}
U=
\begin{pmatrix}
U_0 & 0  \\
0 & U_1
\end{pmatrix},
\label{umatrix}
\end{equation}
where $U_0$ and $U_1$ are unitary matrices which turn $m_{light}$ and  $m_{heavy}$, respectively, diagonal \footnote{Any negative mass eigenvalue can be turned in a positive one by defining a diagonal matrix $K$ such that $UK$ leads to a diagonal form for Eq.~(\ref{dmatrix})  with all entries nonnegative, but it can be omitted without further consequences.}. Thus, the matrix which diagonalizes $M_\nu$ is then,
\begin{equation}
{\cal U}=W\,U=
\begin{pmatrix}
\sqrt{1+FF^\dagger}U_0 & F\,U_1  \\
F^\dagger\,U_0 & \sqrt{1+F^\dagger F}U_1
\end{pmatrix}.
\label{wumatrix}
\end{equation}
Let ${\bf n}_L$ be the $9\times 1$ vector whose components are neutrino mass eigenstates, where we denote ${\bf n}_{iL}^0$, $a=1,2,3$ being the three light mass eigenstates and ${\bf n}_{kL}^1$, $k=1,...,6$, the six heavy ones, so that,
\begin{equation}
{\bf n}_L=
\begin{pmatrix}
{\bf n}_{L}^0  \\
{\bf n}_{L}^1
\end{pmatrix}
={\cal U}^\dagger
\begin{pmatrix}
\nu_{L} \\
s_{L}
\end{pmatrix},
\label{n}
\end{equation}
where $s_{L}=[\nu^C_{L}\,\,N^C_L]^T$ is a $6\times 1$ vector. The flavor eigenstates $\nu_{aL}$ figuring in charged current  are given by the following superposition,
\begin{eqnarray}
\nu_{aL} \approx \left[U_0-\frac{1}{2}F_1F^\dagger_1U_0\right]_{ai}{\bf n}_{iL}^0+(F_1\,U_1)_{ak}{\bf n}_{kL}^1.
\label{nul}
\end{eqnarray}
The matrix connecting the flavor and light mass eigenstates is given by,
\begin{equation}
{\cal N}=(1-\eta)U_0,
\label{npmns}
\end{equation}
where $\eta$ is defined as $\eta \equiv \frac{1}{2}F_1F_1^{\dagger}$. ${\cal N}$ is  nonunitary and replaces the unitary PMNS matrix which parametrizes the mixing in the typical three neutrino scenario. The PMNS matrix is to be identified here with  $U_0$ in Eq. (\ref{umatrix}). All the nonunitarity effects are characterized by $\eta$ which is approximately given by,
\begin{equation}
\eta\equiv\frac{1}{2}F_1F^{\dagger}_1\approx \frac{1}{2}m_D^\dagger (M^{-1})^* (M^{-1})^T m_D.
\label{eta}
\end{equation}

Observation of oscillation phenomena involves charged currents interactions. Since the three left-handed neutrinos  entering in such interactions are now  a superposition of the nine mass eigenstates, as given by Eq.~(\ref{nul}), we have the following charged current Lagrangian,
\begin{eqnarray}
{\cal L}_{CC}&=&-\frac{g}{\sqrt{2}}\bar l_{aL} \gamma^\mu\nu_{aL} W^-_\mu +H.c.
\nonumber\\
&\approx &-\frac{g}{\sqrt{2}}\bar l_{aL} \gamma^\mu\left\{{\cal N}_{ai}{\bf n}_{iL}^0+{\cal K}_{ak}{\bf n}_{kL}^1\right\}W^-_\mu +H.c,
\label{CC}
\end{eqnarray}
where ${\cal K}_{ak}=(F_1\,U_1)_{ak}$. For the term in Eq.~(\ref{CC}) involving the heavy neutrinos ${\bf n}_{kL}^1$ there is a suppression coming  from elements of $(F_1\,U_1)_{ak}$ which are expected at least of order $v_\rho/v_\chi$. It can be estimated, by taking  $v_\rho\approx 10^2$ GeV and $v_\chi\approx 10^3$ GeV, to be of order 10$^{-1}$. This gives a sizable mixing among left-handed and right-handed neutrinos which can be probed through rare lepton flavor violating processes, that we are going to address in a moment.

Returning to $m_{light}$,  on substituting  $m_D=Gv_\rho$, $M=G^{\prime} v_{\chi^{\prime}}$, we obtain
\begin{equation}
m_{light} =\left( G^T (G^{ \prime T})^{-1}\mu (G^{ \prime})^{ -1} G\right)\frac{v^2_\rho }{v^2_{\chi^{\prime}}}.
\label{inverseseesaw331}
\end{equation}

Remember that  $G$ is an anti-symmetric matrix, implying that one eigenvalue of the neutrino mass matrix in Eq.~(\ref{inverseseesaw331}) is null. Then, automatically the squared mass difference in  Eq.~(\ref{currentneutrinodata}) provides, necessarily,  the following neutrino mass spectrum,
\begin{equation}
m_1=0\,\,\,,\,\,\,m_2 \approx 4.8\times 10^{-2}\mbox{eV}\,\,\,,\,\,\,m_3 \approx  8.7 \times 10^{-3}\mbox{eV}.
\label{eigenvalues1}
\end{equation}
In other words, in the framework of the ISS mechanism developed here, the solar and atmospheric neutrino oscillation experiments provide the absolute mass of the neutrinos.

In view of this, let us check if  $m_\nu$ in Eq.~(\ref{inverseseesaw331})  is capable of providing the mass spectrum given in Eq.~(\ref{eigenvalues1}) and the correct mixing matrix. For this  we have to  diagonalize $m_\nu$  in Eq.~(\ref{inverseseesaw331}). However, notice that it involves many free parameters in the form of Yukawa couplings, the components of $G$ and $ G^{\prime}$ and, unfortunately, there are no constraints over them.

With so large set of free parameters, there is a great deal of possible solutions that lead to the correct neutrino mass spectrum and mixing. However due to  the non-unitarity of the mixing matrix ${\cal N}$, any  set of values for the entries in $G$ and $G^{\prime}$ that do the job must obey the following constraints~\cite{nonunitarity},
\begin{equation}
|\eta|<
\begin{pmatrix}
2.0\times 10^{-3} & 3.5\times 10^{-5}  & 8.0\times 10^{-3}  \\
3.5\times 10^{-5} & 8.0\times 10^{-4}  & 5.1\times 10^{-3} \\
8.0\times 10^{-3}  & 5.1\times 10^{-3}  & 2.7\times 10^{-3}
\end{pmatrix}.
\label{nonunitatitybounds}
\end{equation}

In what concerns the energy parameters that appear in $m_\nu$ above, we consider  $v_\eta=v_\rho=v$, thus the constraint  $v^2_\eta + v^2_\rho=(246\mbox{GeV})^2$ imposes  $v=174$GeV. The natural value of $v_{\chi^{\prime}}$ is around TeV, but since we are interested only in its order of magnitude, we consider  exactly $1$~TeV. On the other hand, the usual scale of $\mu$ is around keV. Here we consider   $\mu=0.3\,{\bf{I}}$ keV where ${\bf{I}}$ is the identity matrix.

Regarding the Yukawa couplings entries in $G$ and $G^{\prime}$, notice that $G$ is anti-symmetric matrix. Thus it has only three independent free parameters. In view of this we cannot make the common assumption of considering $M$ as being diagonal and degenerate mass matrix once we are at risk of having less free parameters than necessary to give the correct pattern of neutrino masses and mixing. Thus the simplest scenario here is one  where $M$ is diagonal but non-degenerate. But even in this case it is not possible to uniquely fix the parameters in $G$ and $G^{\prime}$. In what follows we present a particular  solution of the diagonalization of the mass matrix $m_\nu$ above which involves the following set of values for $G$ and $G^{\prime}$ entries,
\begin{equation}
G=
\begin{pmatrix}
0 & 0.02 & 0.012 \\
-0.02& 0.0 & 0.01\\
-0.012 & \,\,\,\,\,-0.01 & 0.0
\end{pmatrix}\,\,\,,\,\,\,G^{\prime}=
\begin{pmatrix}
0.32& 0.0 & 0.0 \\
0.0& 0.8 & 0.0\\
0.0 & \,\,\,\,\,0.0 & 0.9
\end{pmatrix}.
\label{parametersvalues}
\end{equation}
With these  $G$, $G^{\prime}$ and the values for the  VEV's $v$, $v_{\chi^{\prime}}$ and $\mu$ presented above, the diagonalization of the mass matrix $m_{light}$ in Eq. (\ref{inverseseesaw331}) furnishes the desired eigenvalues in Eq.~(\ref{eigenvalues1}) and, in addition,  yields a standard PMNS mixing matrix  given by,
\begin{equation}
U_{PMNS}=
\begin{pmatrix}
0.802987 & 0.583404  & 0.121869  \\
-0.485344 & 0.521409  & 0.701836 \\
0.34591  & -0.622714  & 0.701836
\end{pmatrix}.
\label{PMNSprediction}
\end{equation}
This $U_{PMNS}$ implies in the following mixing angles $\theta_{12} = 36^o$, $\theta_{23} =
 45^o$ and   $\theta_{13} = 7^o$.

This set of values for the entries in $G$ and $G^{\prime}$ yields,
\begin{equation}
|\eta|=
\begin{pmatrix}
1.4\times 10^{-5} & -5.0\times 10^{-10}  & 4.7\times 10^{-6}  \\
-5.0\times 10^{-10} & 3.6\times 10^{-5}  & 3.9\times 10^{-5} \\
4.7\times 10^{-6}  & 3.8\times 10^{-5}  & 4.2\times 10^{-5}
\end{pmatrix},
\label{nonunitatityprediction}
\end{equation}
which  respect the bounds in Eq.~(\ref{nonunitatitybounds}).

Let us see the prediction for the  six right-handed neutrinos masses  that such set of values for the parameters in $G$ and $G^{\prime}$ can provide.  On diagonalizing $m_{heavy}={\cal M}_{R}$ in Eq.~(\ref{definition}), we obtain two eigenvalues around $900$~GeV, other two around $800$~GeV and two more around $320$~GeV. With these masses such heavy neutrinos  may be probed in the LHC through the process $pp\rightarrow l^{\pm}l^{\pm}l^{\mp}\nu(\bar \nu)$~\cite{probedLHC}, or in future neutrino  experiments through rare lepton decays, like $\mu \rightarrow e \gamma$.

We focus  now on  the rare   lepton flavor violation(LFV) process $\mu \rightarrow e\gamma$. Such process is allowed by the second coupling in Eq.~(\ref{CC}). The branching ratio for the process mediated by these  six heavy neutrinos is given by~\cite{pilaftis},
\begin{equation}
BR(\mu \rightarrow e \gamma)\approx \frac{\alpha^3_W \sin^2(\theta_W) m^5_{\mu}}{256 \pi^2 m^4_W \Gamma_\mu}\times |\sum_{1=1}^6 {\cal K}_{\alpha i}{\cal K}_{\beta i}I(\frac{m^2_{N_i}}{m^2_W})|^2,
\label{BR}
\end{equation}
where
\begin{equation}
I(x)=-\frac{2x^3+5x^2-x}{4(1-x)^3}-\frac{3x^3\ln x}{2(1-x)^4}.
\end{equation}

In the above branching ratio $\alpha_W= \frac{g^2}{4 \pi}$ with $g$ being the weak coupling, $\theta_W$ is the  electroweak mixing angle, $m_\mu$ is the muon mass, $m_W$ is the $W^{\pm}$ mass, $\Gamma_\mu$ is the muon total decay width. The present values of these parameters are found in Ref.~\cite{PDG}.  In order to obtain ${\cal K}$, we get $F_1$ from Eq.~(\ref{mlmh}) and diagonalize ${\cal M}_{R}$ in  Eq.~(\ref{definition})  to obtain $U_1$.

Considering all this, we obtain the approximate value for $BR(\mu \rightarrow e \gamma)$,
\begin{equation}
BR(\mu \rightarrow e \gamma)\approx 3\times 10^{-14}.
\label{BRvalue}
\end{equation}
The current upper bound on this branching ratio is $BR(\mu \rightarrow e \gamma)< 4.9\times 10^{-11}$\cite{PDG}. Our result for this branching ratio respects the  upper bound and, interestingly, falls inside the sensitivity of  future neutrino experiments~\cite{futureexperiments}, which will be able to probe a branching ratio up to $10^{-18}$, representing an additional test of our proposal.

\section{A possible realization of a small $\mu$}
In this section we develop a dynamical explanation to the smallness of the parameter $\mu$. Basically we adapt the mechanism developed in Ref.~\cite{explanation1} to our context.

We first remark that the 331RHN model with only an additional discrete symmetry does not work in the appropriate way as to furnish the correct entries in the mass matrix Eq.~(\ref{ISSmatrix}). In view of this we thought of a minimal modification of this model and added a scalar singlet $\sigma\sim(1\,,\,0)$ to its scalar content in Eq.~(\ref{scalarcontent}). In order to avoid unpleasant terms in the Lagrangian of the model, we can then impose a $Z_3$ symmetry  with only the following fields transforming nontrivially according to the following assignement,
\begin{equation}
N_{aR} \rightarrow e^{2i \pi/3}N_{aR}\,,\,\,L_a =e^{4i \pi/3}L_a\,,\,\,\,\sigma \rightarrow e^{2i \pi/3} \sigma \,,\,\, \chi \rightarrow e^{2i \pi/3} \chi\,,\,\,\, \rho=e^{4i \pi/3}\rho.
\label{yukawa331}
\end{equation}
This discrete symmetry has a twofold importance. It restricts the Yukawa interaction terms leading, after spontaneous symmetry breaking, to mass terms needed for producing the texture like that in Eq.~(\ref{ISSmatrix}). Also, the $Z_3$ symmetry plays a role in the scalar potential allowing terms, like the trilinear one, which guarantees a safe spectrum of scalar fields (meaning, no extremely light scalar).

The Yukawa Lagrangian of interest to the implementation of the ISS mechanism involving scalars and leptons, invariant by $Z_3$, is composed by the following sum of terms,
\begin{equation}
{\cal L}_{Y} = g_{ab}\epsilon_{ijk}\overline{(L_{a_i})^c} L_{b_j}\rho_k + g^{\prime}_{ab}\overline{L_a} \chi N_{b_R} + \frac{\lambda_{ab}}{2}\sigma^0 \overline{N^c_{aL}} N_{bR} + H.c.
\label{newYI}
\end{equation}

Let us assume the following shift on the neutral scalars,
\begin{equation}
\eta^{0}, \eta^{\prime}, \rho^{0}, \chi^{\prime}, \sigma^0 \rightarrow \frac{1}{\sqrt{2}} (v_{\eta, \eta^{\prime}, \rho, \chi^{\prime}, \sigma} + R_{\eta, \eta^{\prime}, \rho, \chi^{\prime}, \sigma} + iI_{\eta, \eta^{\prime}, \rho, \chi^{\prime}, \sigma}).
\label{scalarshift}
\end{equation}
With these  VEVs  the Yukawa terms in Eq.~(\ref{newYI}) provide the  neutrino mass terms in Eq.~(\ref{massterms}) with $\mu$  being recognized as $\mu=\frac{\lambda v_\sigma}{\sqrt{2}}$. In this case a small $\mu$ requires a small $v_\sigma$. In order to achieve this we evoke a kind of type II seesaw mechanism~\cite{seesawII} over $v_\sigma$, built from the scalar  potential that obeys the extra $Z_3$ symmetry,
\begin{eqnarray}
V & = & (\mu_{1}^{2}+\lambda_{1}|\eta|^{2})|\eta|^{2}+(\mu_{2}^{2}+\lambda_{2}|\rho|^{2})|\rho|^{2}+
(\mu_{3}^{2}+\lambda_{3}|\chi|^{2})|\chi|^{2}+(\mu^2_4 +\lambda_4|\sigma|^2) |\sigma|^2\nonumber\\
  & + & \lambda_{5}|\eta|^{2}|\rho|^{2}+
  \lambda_{6}|\eta|^{2}|\chi|^{2}+
  \lambda_{7}|\rho|^{2}|\chi|^{2}+ \lambda_{8}|\eta^{\dagger}\rho|^2 +
  \lambda_{9}|\eta^{\dagger}\chi|^2 +
  \lambda_{10}|\rho^{\dagger}\chi|^2\nonumber\\
  & + &  (\lambda_{11}|\eta|^{2}
  +\lambda_{12}|\rho|^{2}+\lambda_{13}|\chi|^{2})|\sigma|^2  +
  \left(\frac{f_1}{\sqrt{2}}\epsilon^{ijk}\eta_{i}\rho_{j}\chi_{k}+\frac{f_2}{\sqrt{2}}\chi^{\dagger}\eta \sigma+\frac{f_3}{\sqrt{2}}\sigma^3+H.c.\right).
  \label{Vt}
\end{eqnarray}

The existence of a minimum of the potential  in Eq.~(\ref{Vt})  requires its first derivatives, with respect to the neutral scalar fields developing VEV, to vanish, which leads to a set of five constraint equations . However,  for our proposal here the only constraint equation that matters is the one related to the neutral scalar field  $\sigma^0$,
\begin{eqnarray}
v_\sigma[2\mu^2_\sigma + 2\lambda_4 v^2_\sigma + \lambda_{11}(v^2_\eta+v^2_{\eta^{\prime}})  + \lambda_{12}v^2_\rho + \lambda_{13}v^2_{\chi^{\prime}}+3f_3 v_\sigma]+f_2 v_\chi v_{\eta^{\prime}} =0.
\label{constraint331}
\end{eqnarray}

The traditional assumption here is that the scalar $\sigma^0$ is very heavy belonging to a GUT scale~\cite{seesawII}.  In this case on assuming that $\mu_\sigma $ is the dominant energy parameter in the constraint equation above, we obtain,
\begin{equation}
v_\sigma\approx \frac{f_2 v_{\eta^{\prime}}  v_{\chi^{\prime}}}{\mu^2_\sigma}.
\label{331vev}
\end{equation}
As $f_2$ is related to a term that explicitly violates lepton number, it is also usual to assume that it belongs to the GUT scale, too. Assuming that the GUT scale is $\Lambda$ we must have $\Lambda=\mu_\sigma = f_2$. In this case we obtain,
\begin{equation}
v_\sigma\approx \frac{ v_{\eta^{\prime}}  v_{\chi^{\prime}}}{\Lambda}.
\label{331ss}
\end{equation}

There is an upper bound over $v_{\eta^{\prime}}<40$~GeV derived in Ref.~\cite{upperbounds}. On assuming reasonable values for the VEVs of the model, $v_{\eta^{\prime}}=10$~GeV, $v_{\chi^{\prime}}=10^3$~GeV and $\Lambda=10^{10-11}$~GeV, we obtain $v_\sigma\approx 0.1-1$KeV, which implies  $\mu$ around KeV.

\section{conclusions}

The appealing point behinds the ISS mechanism  is the fact that it is a phenomenological seesaw mechanism whose signature are right-handed neutrinos at TeV scale which may be probed at LHC and in future neutrino experiments through rare LFV process.

Although the  ISS mechanism is a phenomenologically feasible seesaw mechanism, it lacks a natural underlying framework, namely one that accommodates right-handed neutrinos at TeV scale. In this work we  implemented the ISS mechanism in the 331RHN~\footnote{We remark that when we concluded this manuscript a similar attempt was proposed in Ref.~\cite{ochoa}.}. We showed that the model possesses the appropriate neutrino content  and  the energy scales as required by the mechanism. We also provided a concrete example which recovered the neutrino physics involved in oscillation neutrino experiments, and evaluated the rare LFV decay $\mu \rightarrow e\gamma$ whose prediction is around $BR(\mu \rightarrow e \gamma)\approx 3\times 10^{-14}$. Such robust value for this branching ratio may be probed at future neutrino experiments and represents a further means of testing our proposal.  Finally, we developed a scheme where the model can be suitably modified to provide a natural explanation of the smallness of the characteristic ISS parameter $\mu$ (keV scale).

In view of all these results, it seems that the 3331RHN model is an interesting framework for realizing the ISS mechanism.

%%%%%%%%%%%%%%%%%%%%%%%%%%%%%%%%%%%%%%%%%%%%%%%%%%%%%%%%%%%%%%%%%%%%%%%%%%%%%%%%%%%%%%%%%%%
\acknowledgments
This work was supported by Conselho Nacional de Pesquisa e
Desenvolvimento Cient\'{i}fico- CNPq (C.A.S.P, P.S.R.S and A.G.D) and Coordena\c c\~ao de Aperfei\c coamento de Pessoal de N\'{i}vel Superior - CAPES (A.S). A.G.D. also thanks FAPESP for supporting this work.
%%%%%%%%%%%%%%%%%%%%%%%%%%%%%%%%%%%%%%%%%%%%%%

%%%%%%%%%%%%%%%%%%%%%%%%%%%%%%%%%%%%%%%%%%%%%%%%%%%%%%%%%%%%%%%%%%%%%%%%%%%%%%%%%%%%%%%%%%%

\end{document}